\documentclass{lmcs} 
\pdfoutput=1

\usepackage{lastpage}
\lmcsdoi{17}{2}{25}
\lmcsheading{}{\pageref{LastPage}}{}{}%
{Oct.~19,~2018}{Jun.~29,~2021}{}

\usepackage{amsfonts,amsthm,amssymb,amsmath,color}

\usepackage[utf8]{inputenc}

\usepackage{hyperref}

\newdimen\proofrulebreadth \proofrulebreadth=.05em
\newdimen\proofdotseparation \proofdotseparation=1.25ex
\newdimen\proofrulebaseline \proofrulebaseline=2ex
\newcount\proofdotnumber \proofdotnumber=3
\let\then\relax
\def\hfi{\hskip0pt plus.0001fil}
\mathchardef\squigto="3A3B
\newif\ifinsideprooftree\insideprooftreefalse
\newif\ifonleftofproofrule\onleftofproofrulefalse
\newif\ifproofdots\proofdotsfalse
\newif\ifdoubleproof\doubleprooffalse
\let\wereinproofbit\relax
\newdimen\shortenproofleft
\newdimen\shortenproofright
\newdimen\proofbelowshift
\newbox\proofabove
\newbox\proofbelow
\newbox\proofrulename
\def\shiftproofbelow{\let\next\relax\afterassignment\setshiftproofbelow\dimen0 }
\def\shiftproofbelowneg{\def\next{\multiply\dimen0 by-1 }%
\afterassignment\setshiftproofbelow\dimen0 }
\def\setshiftproofbelow{\next\proofbelowshift=\dimen0 }
\def\setproofrulebreadth{\proofrulebreadth}

\def\prooftree{
\ifnum  \lastpenalty=1
\then   \unpenalty
\else   \onleftofproofrulefalse
\fi
\ifonleftofproofrule
\else   \ifinsideprooftree
        \then   \hskip.5em plus1fil
        \fi
\fi
\bgroup
\setbox\proofbelow=\hbox{}\setbox\proofrulename=\hbox{}%
\let\justifies\proofover\let\leadsto\proofoverdots\let\Justifies\proofoverdbl
\let\using\proofusing\let\[\prooftree
\ifinsideprooftree\let\]\endprooftree\fi
\proofdotsfalse\doubleprooffalse
\let\thickness\setproofrulebreadth
\let\shiftright\shiftproofbelow \let\shift\shiftproofbelow
\let\shiftleft\shiftproofbelowneg
\let\ifwasinsideprooftree\ifinsideprooftree
\insideprooftreetrue
\setbox\proofabove=\hbox\bgroup$\displaystyle 
\let\wereinproofbit\prooftree
\shortenproofleft=0pt \shortenproofright=0pt \proofbelowshift=0pt
\onleftofproofruletrue\penalty1
}

\def\eproofbit{
\ifx    \wereinproofbit\prooftree
\then   \ifcase \lastpenalty
        \then   \shortenproofright=0pt  
        \or     \unpenalty\hfil         
        \or     \unpenalty\unskip       
        \else   \shortenproofright=0pt  
        \fi
\fi
\global\dimen0=\shortenproofleft
\global\dimen1=\shortenproofright
\global\dimen2=\proofrulebreadth
\global\dimen3=\proofbelowshift
\global\dimen4=\proofdotseparation
\global\count255=\proofdotnumber
$\egroup  
\shortenproofleft=\dimen0
\shortenproofright=\dimen1
\proofrulebreadth=\dimen2
\proofbelowshift=\dimen3
\proofdotseparation=\dimen4
\proofdotnumber=\count255
}

\def\proofover{
\eproofbit 
\setbox\proofbelow=\hbox\bgroup 
\let\wereinproofbit\proofover
$\displaystyle
}%
\def\proofoverdbl{
\eproofbit 
\doubleprooftrue
\setbox\proofbelow=\hbox\bgroup 
\let\wereinproofbit\proofoverdbl
$\displaystyle
}%
\def\proofoverdots{
\eproofbit 
\proofdotstrue
\setbox\proofbelow=\hbox\bgroup 
\let\wereinproofbit\proofoverdots
$\displaystyle
}%
\def\proofusing{
\eproofbit 
\setbox\proofrulename=\hbox\bgroup 
\let\wereinproofbit\proofusing
\kern0.3em$
}

\def\endprooftree{
\eproofbit 
  \dimen5 =0pt
\dimen0=\wd\proofabove \advance\dimen0-\shortenproofleft
\advance\dimen0-\shortenproofright
\dimen1=.5\dimen0 \advance\dimen1-.5\wd\proofbelow
\dimen4=\dimen1
\advance\dimen1\proofbelowshift \advance\dimen4-\proofbelowshift
\ifdim  \dimen1<0pt
\then   \advance\shortenproofleft\dimen1
        \advance\dimen0-\dimen1
        \dimen1=0pt
        \ifdim  \shortenproofleft<0pt
        \then   \setbox\proofabove=\hbox{%
                        \kern-\shortenproofleft\unhbox\proofabove}%
                \shortenproofleft=0pt
        \fi
\fi
\ifdim  \dimen4<0pt
\then   \advance\shortenproofright\dimen4
        \advance\dimen0-\dimen4
        \dimen4=0pt
\fi
\ifdim  \shortenproofright<\wd\proofrulename
\then   \shortenproofright=\wd\proofrulename
\fi
\dimen2=\shortenproofleft \advance\dimen2 by\dimen1
\dimen3=\shortenproofright\advance\dimen3 by\dimen4
\ifproofdots
\then
        \dimen6=\shortenproofleft \advance\dimen6 .5\dimen0
        \setbox1=\vbox to\proofdotseparation{\vss\hbox{$\cdot$}\vss}%
        \setbox0=\hbox{%
                \advance\dimen6-.5\wd1
                \kern\dimen6
                $\vcenter to\proofdotnumber\proofdotseparation
                        {\leaders\box1\vfill}$%
                \unhbox\proofrulename}%
\else   \dimen6=\fontdimen22\the\textfont2 
        \dimen7=\dimen6
        \advance\dimen6by.5\proofrulebreadth
        \advance\dimen7by-.5\proofrulebreadth
        \setbox0=\hbox{%
                \kern\shortenproofleft
                \ifdoubleproof
                \then   \hbox to\dimen0{%
                        $\mathsurround0pt\mathord=\mkern-6mu%
                        \cleaders\hbox{$\mkern-2mu=\mkern-2mu$}\hfill
                        \mkern-6mu\mathord=$}%
                \else   \vrule height\dimen6 depth-\dimen7 width\dimen0
                \fi
                \unhbox\proofrulename}%
        \ht0=\dimen6 \dp0=-\dimen7
\fi
\let\doll\relax
\ifwasinsideprooftree
\then   \let\VBOX\vbox
\else   \ifmmode\else$\let\doll=$\fi
        \let\VBOX\vcenter
\fi
\VBOX   {\baselineskip\proofrulebaseline \lineskip.2ex
        \expandafter\lineskiplimit\ifproofdots0ex\else-0.6ex\fi
        \hbox   spread\dimen5   {\hfi\unhbox\proofabove\hfi}%
        \hbox{\box0}%
        \hbox   {\kern\dimen2 \box\proofbelow}}\doll%
\global\dimen2=\dimen2
\global\dimen3=\dimen3
\egroup 
\ifonleftofproofrule
\then   \shortenproofleft=\dimen2
\fi
\shortenproofright=\dimen3
\onleftofproofrulefalse
\ifinsideprooftree
\then   \hskip.5em plus 1fil \penalty2
\fi
}

\input xy
  \xyoption{all}
  \xyoption{color}
\newdir{>>>}{!/2pt/\dir{>>}*!/6.7pt/\dir{-}*!/6.7pt/\dir{>}}
\newdir{<<<}{!/2pt/\dir{<<}*!/6.7pt/\dir{-}*!/6.7pt/\dir{<}}
   \CompileMatrices
   \UseComputerModernTips
   
   \everyentry={\vphantom(}

\title[Encoding many-valued logic in $\lambda$-calculus]{Encoding many-valued logic in $\lambda$-calculus}
\author[F.J. de Vries]{Fer-Jan de Vries}	
\address{Informatics, University of Leicester, UK}	
\email{fer.jan.de.vries@gmail.com}  

\keywords{lambda calculus, many-valued logic, meaningless term, generalised B{\"o}hm tree, Russell's paradox}

\newcommand{\ProofRule}[3]{
\prooftree
   {#1}  
\justifies 
   {#2}
 \using (#3)
\endprooftree
}

\newcommand{\leftcircwedge}
      {\hbox{\lower.65ex\hbox {{\scriptsize$ \circ$}}}{\kern-.205em \hbox{$\wedge$}}}
\newcommand{\leftcircvee}
      {\hbox{\raise1.3ex\hbox{{\scriptsize$ \circ$}}}{\kern-.215em \hbox{$\vee$}}}
\newcommand{\lsand}{{\leftcircwedge}}
\newcommand{\lsor} {{\leftcircvee  }}

\def\eg{{\em e.g.}}

\newbox\gnBoxA
\newdimen\gnCornerHgt
\setbox\gnBoxA=\hbox{$\ulcorner$}
\global\gnCornerHgt=\ht\gnBoxA
\newdimen\gnArgHgt
\def\Quinequote #1{%
    \setbox\gnBoxA=\hbox{$#1$}%
    \gnArgHgt=\ht\gnBoxA%
    \ifnum     \gnArgHgt<\gnCornerHgt \gnArgHgt=0pt%
    \else \advance \gnArgHgt by -\gnCornerHgt%
    \fi \raise\gnArgHgt\hbox{$\ulcorner$} \box\gnBoxA %
    \raise\gnArgHgt\hbox{$\urcorner$}}

\newcommand{\Uset}{{\mathcal{U}}}

\newcommand{\False}{{\mathbf{F}}}

\newcommand{\True}{{\mathbf{T}}}

\newcommand{\limp}{{\rightarrow}}
\newcommand{\ITE}[3]{\mathbf{if\ }{#1}\ \mathbf{then\ }{#2}\ \mathbf{else\ }{#3}}

\usepackage{colordvi}
\usepackage{color}
\usepackage[curve,matrix,arrow,cmtip]{xy} \CompileMatrices
\UseComputerModernTips  \everyentry={\vphantom(}
\usepackage{cancel}

 \def \bchpaula {\begin{color}{blue}} 
 \def \echpaula {\end{color}}

 \def \bchferjan {\begin{color}{red}} 
 \def \echferjan {\end{color}}
 
\renewcommand{\next}{{\sf next\ }}

\newcommand{\Kterm}{{\mathbf K}}

\newcommand{\Oterm}{{\mathbf O}}
\newcommand{\omegaterm}{{\mathbf \Omega}}
\newcommand{\Omegaterm}{{\mathbf \Omega}}

\makeatletter
\newcommand{\dotminus}{\mathbin{\text{\@dotminus}}}

\newcommand{\@dotminus}{%
  \ooalign{\hidewidth\raise1ex\hbox{.}\hidewidth\cr$\m@th-$\cr}%
}
\makeatother

\newcommand{\Iterm}{\mathbf{I}}
\newcommand{\Yterm}{\mathbf{\Theta}}

\def \one  {\mathrel{\rightarrow}}
\def \fin  {\mathrel{\rightarrow\!\!\!\!\!\rightarrow}}
\def \finstep  {\fin}
\def\many{\mathrel{\rightarrow\!\!\!\!\!\rightarrow\!\!\!\!\!\rightarrow}}

\def \manystep{\many}

\newcommand{\finbeta}{\fin_{\beta}}
\newcommand{\manybeta}{\many_{\beta}}

\newcommand{\setHA}{\mathcal {HA}}
\newcommand{\setIL}{\mathcal {IL}}

\newcommand{\RA}{\setR}

\newcommand{\setO}{{\mathcal O}}

\newcommand{\Lbi}{{\sf \Lambda}_{\bot}^\infty}

\newcommand{\Li}{{\sf \Lambda}^\infty}

\newcommand{\Linob}{{\sf \Lambda}^\infty}

\newcommand{\Lib}{\Lbi}

\newcommand{\setUnsolvables}{\mathcal US}
\newcommand{\setU}{\mathcal U}

\newcommand{\setR}{\setroot}
\newcommand{\setogre}{{\mathcal O}}

\newcommand{\setroot}{{\mathcal R}}

\newcommand{\sethr}{{\mathcal H A}}

\newcommand{\botHA}
           {\bot_\setHA}
\newcommand{\botIL}
           {\bot_\setIL}
\newcommand{\botO}
           {\bot_\setO}
\newcommand{\botILO}{\bot_{\setIL\cup\setO}}

\newcommand{\setil}{{\mathcal I L}}

\newcommand{\indis}{\stackrel{\setU}{\leftrightarrow}}

\newcommand{\lb}{\lambda_{\beta}}

\newcommand{\li}{\lambda^{\infty}}

\newcommand{\libBU}{\libbU{\setU}}
\newcommand{\libBohm}{\libbU{\setUnsolvables}}

\newcommand{\libbU}[1]{\li_{\beta {\bot_{#1}}}}
\newcommand{\libbUU}[2]{\li_{\beta {\bot_{#1}} {\bot_{#2}} }}
\newcommand{\libbUUU}[3]{\li_{\beta {\bot_{#1}} {\bot_{#2}} {\bot_{#3}}}}

\begin{document}
\bibliographystyle{alpha}
\maketitle

\pdfbookmark[0]{Dedication}{dedication} 

\begin{abstract}
We will extend the well-known Church encoding of Boolean logic
into $\lambda$-calculus to an encoding of  McCarthy's $3$-valued logic  into a suitable
infinitary extension of $\lambda$-calculus that identifies all unsolvables by $\bot$, where $\bot$ is a fresh constant. 
This encoding refines to $n$-valued logic, for $n\in\{4,5\}$. Such encodings also exist for Church's original $\lambda\Iterm$-calculus.

By way of motivation  we consider Russell's paradox, exploiting the fact that the same encoding allows us  also to calculate  truth values of infinite closed propositions  in this infinitary setting.

\end{abstract}

\section*{In memory of Corrado B\"ohm}  B\"ohm's theorem~\cite{Bohm1968}
was instrumental in proving the equivalence between an operational semantics and
a denotational semantics of the $\lambda$-calculus and inspired Barendregt~\cite{bare77,Barendregt1984} to
define the concept of B\"ohm tree, a first version of which had been introduced
by B\"ohm and Dezani~\cite{BohmDezani1974}. B\"ohm trees have later
 been redefined as the normal forms in a suitable infinitary extension of
$\lambda$-calculus by Kennaway et al.~\cite{KKSV97}. B\"ohm trees and their generalisations are
now another established way to capture the semantic content of
a $\lambda$-term~\cite{KV03,SV11}. In this paper B\"ohm trees play a crucial role: we use B\"ohm trees to encode (even infinite) propositions in $\lambda$-calculus
and to calculate their values.

\section{Motivation and overview}
In this paper we will extend the well-known Church encoding of Boolean logic
into $\lambda$-calculus to an encoding of $n$-valued logic (for $ 3 \leq n \leq 5$) into an appropriate
infinitary extension of $\lambda$-calculus. The extension we use in case of $n=3$ is the extension that identifies all unsolvables by $\bot$ such that the normal forms of the lambda terms are their B\"ohm trees. 
By way of motivation  we will now consider Russell's paradox. Any notation that
is used in this section  will be explained in Section~\ref{sec2} and~\ref{sec3}.
\subsection{Russell's Paradox}
This paradox arises if we, somewhat na\"ively, consider the set $R$ of
  all sets that are not a member of themselves and then  wonder whether $R \in R$. We  get as   paradoxical consequence that $R \in R$ if and only if $R \notin R$. 
 As noted by Church~\cite{Church1941} at the heart of this paradox lies the  $\lambda$-term (the application $p p$ is interpreted as $p \ni p$)
 \[P \equiv (\lambda p . \lnot(p p))(\lambda p . \lnot(p
   p))\] which has no finite normal form and therefore is neither true or false.\footnote{The notation $\lnot$ for the $\lambda$-term $\lambda b.b\True\False$ is recalled at page~\pageref{lnot}.}  Like the well-known $\lambda$-term
  $\Omegaterm \equiv (\lambda x . x x)(\lambda x .x x)$, the term $P$ has the property that any of its reducts can be further reduced
   to a redex. Therefore it has no head normal form. Hence $P$ is an
  unsolvable term. It allows  the following infinite reduction
\[P \to \neg P\to \neg (\neg P)\to \neg (\neg (\neg P)) \to \dots \]
The limit $\lnot(\neg (\neg (\neg (\ldots))))$ of this reduction is an 
infinite proposition. Unexpected, perhaps, but not necessarily paradoxical. 
\subsection{Infinite $\lambda$-calculus and B\"ohm trees}
In the past~\cite{KV03,SV11} we have developed a family of infinitary $\lambda$-calculi, each
depending on a set of meaningless terms $\Uset$. The set of terms underlying these extensions is the set $\Lib$ of lambda terms obtained by interpreting the usual $\lambda$-calculus syntax extended with one fresh symbol $\bot$ coinductively. We use this set $\Uset \subseteq \Lib$ of meaningless terms to add a new rewrite rule to  $\lb$ that  allows us to rewrite meaningless terms  to $\bot$. Cf.\ Section~\ref{infinitelambdacalculus} for a precise description of this rule.

The set $\setUnsolvables$ of unsolvable $\lambda$-terms is
the best known example of such a set of meaningless terms.
The corresponding infinitary extension $\libBohm$ of the
  finite $\lambda$-calculus $\lb$ is
  confluent and normalising for a suitable notion of possibly infinite
  reduction. The B\"ohm tree of  a finite $\lambda$-term is precisely its normal form in $\libBohm$.  In particular the B\"ohm tree of an unsolvable is $\bot$.

So with this encoding in the $\lambda$-calculus in mind we no longer need  be afraid of infinite propositions. By inspecting the B\"ohm trees of the encoding of  infinite closed
propositions we will find that they are either lambda terms representing a Boolean or they are unsolvable.
\subsection{Encoding three-valued logic in infinitary $\lambda$-calculus}
 Thus we are led to extend the Church encoding to an encoding of three-valued logic in
infinitary $\lambda$-calculus $\libBohm$, by mapping the third value to $\bot$. Inspection of
the truth tables then reveals that the Church encoding of Boolean logic now has
naturally been extended to a Church encoding of what is called McCarthy's three-valued logic~\cite{McCarthy1963}.
In particular we find that the infinite term $\neg(\neg(\neg(\ldots)))$ that we
encountered in our analysis of Russell's paradox is neither true nor false but $\bot$.

\subsection{Encoding four- and five-valued logic}
We will further note that the set of unsolvable $\lambda$-terms that get
identified by $\bot$ can be split into three subsets closed under infinite
reduction and substitution. Repeating the above construction now with three
new truth values instead of $\bot$ we find that the Church encoding also encodes a five-valued
McCarthyan logic. That five-valued
logic and its four-valued sub-logic have been studied earlier by
Bergstra and Van de Pol~\cite{BergstraPol1996,BergstraPol2011}.

\subsection{Church's $\lambda\Iterm$-calculus}
When Church started his work on $\lambda$-calculus around or before
1928, his motivation was to use the $\lambda$-calculus as the basis
for a symbolic logic that could serve as the foundation of
mathematics~\cite{Church1932}. Church's hope was that by using
non-classical logic (in which he had shown an early
interest~\cite{Church1928}) he could side step the Paradoxes without
having to introduce Zermelo's set axioms or Russell's type theory,
that he both judged as somewhat artificial.

This is not what happened. He discovered with his
students Kleene and Rosser that the lambda definable functions
corresponded exactly to the recursive
functions~\cite{Kleene1936,Kleene1936b,Church1936Unsolvable}. In the
build-up to that result Kleene and Rosser managed to prove the
inconsistency of his logical system~\cite{KleeneRosser1935} while
Church himself was still publicly hopeful that not only his system could be
paradox-free but also escape G\"odel's incompleteness theorem~\cite{Church1934Richard}.  A disaster. Fortunately, the
$\lambda$-calculus itself was consistent by the Church-Rosser
theorem~\cite{ChurchRosser1936}. Various papers under preparation had
to be rewritten. Church rebounded almost immediately with his
formulation of the Church-Turing thesis~\cite{Church1936Unsolvable}
and his negative solution of Hilbert's {\em
  Entscheidungsproblem}~\cite{Church1936Note} (there is no algorithm
that can decide whether a given formula of the first order arithmetic
is provable or not).

Church's goal, a paradox free system of symbolic logic, led him to the choice of the $\lambda\Iterm$-calculus in which an abstraction $\lambda x.M$ is only accepted as well-formed term if it  contains $x$ as a free variable. For him only terms with a finite normal form where significant and for this he rejected, what we now call, the  classical lambda calculus which has terms that have a normal form although they also have subterms which do not~\cite{Church1941}.

It is a natural question to ask whether an  encoding of $3$-valued logic is possible in the $\lambda\Iterm$-calculus.
  We recall that there is a Church
encoding for Boolean logic in the $\lambda\Iterm$-calculus. Barendregt  has shown
that the unsolvable terms in $\lambda\Iterm$-calculus are exactly the
$\lambda\Iterm$-terms without a finite normal form. This means that the B\"ohm
tree of a $\lambda\Iterm$-term is either its finite normal form or $\bot$. No infinite terms or reductions are needed in case of
$\lambda\Iterm$-calculus to define B\"ohm
trees.  Thus 
the above encoding of  McCarthy's three-valued logic can be quite simply repeated in
Church's $\lambda\Iterm$-calculus. For the details see Section~\ref{lambdaI}.

Yet, while this encoding is undoubtedly well  within Church's technical means, the B\"ohm tree concept seems in conflict with his intuition of meaning. The B\"ohm tree construction gives meaning to any term: the  terms without a finite normal form which Church considers meaningless/insignificant are given the ``meaning'' $\bot$  in this extension of $\lambda\Iterm$-calculus with the $\bot$-rule.as

\subsection{Overview of this paper} In Section~\ref{sec2} we assume familiarity with the finite $\lambda$-calculus and briefly introduce relevant notation and facts from  the infinitary $\lambda$-calculus.
In Section~\ref{sec3} we recall the encoding of Boolean valued logic and explain how to extend this to an encoding of three-valued logic. Then we show how this encoding can be refined to four- and five valued logic.
In Section~\ref{sec4} we discuss B\"ohm trees for Church's $\lambda\Iterm$-calculus and show that three valued logic can also be encoded in $\lambda\Iterm$-calculus.
Finally Section~\ref{sec5} is a brief conclusion.

\section{Infinite $\lambda$-calculus}
\label{sec2}
We will recall notation, concepts and facts from infinitary $\lambda$-calculus,
while assuming familiarity with  $\lb$,  by which we denote the finite
$\lambda$-calculus  with $\beta$-reduction and no
$\eta$-reduction~\cite{Church1941,Barendregt1984}. We will use $\to$ and
$\finstep$ for respectively one step $\beta$-reduction and finite
$\beta$-reduction. We will use $\equiv$ to indicate syntactical identity modulo
$\alpha$. We will use the following special terms.
\[\begin{array}{lclllcl}
\Kterm &\equiv& \lambda xy.y&\quad\quad&\Omegaterm &\equiv& (\lambda x.xx)\,\lambda x.xx\\
\Iterm  &\equiv& \lambda x.x &\quad\quad&\Yterm &\equiv& (\lambda xy.y(xxy))\,\lambda xy.y(xxy)
\end{array}\]

We will now explain how to construct infinite extensions of the finite $\lambda$-calculus that are confluent and normalising. We begin with the observation that finite reduction is not finitely normalising: for instance, the finite term $\Yterm x$ has an infinite reduction
 \[\Yterm x \to x(\Yterm x) \to x(x(\Yterm x)) \to \ldots\]
     This is a converging reduction (think of terms as trees and take the
     standard metric on trees) with an infinite term as limit:
     \[x (x (x(\ldots)))\]
     We can add infinite $\lambda$-terms to the  finite $\lambda$-terms by reading the usual syntax definition (where $x$ ranges over some countable set of variables) of finite $\lambda$-terms coinductively:
\[M ::= x 
\mid \lambda x . M 
\mid (M M)
\]
 We will write $\Li$ for this set of finite and infinite $\lambda$-terms. Using $\manystep$ for a possibly infinite converging reduction, we can now write
\[\Yterm x \manystep x (x( x( \ldots)))\]
Later in the paper we will encounter the infinite term $\lambda y\lambda y\lambda y\ldots$ as the limit of the converging reduction
\[\Yterm \Kterm \to \Kterm (\Yterm \Kterm) \to \lambda y.\Yterm \Kterm\finstep \lambda y \lambda y.\Yterm \Kterm\finstep \lambda y \lambda y \lambda y.\Yterm \Kterm\manystep\lambda y\lambda y\lambda y\ldots\]
These two examples show that by adding infinite terms and infinite reductions to the finite lambda calculus, we obtain that some finite terms without a finite normal form now have converging reductions to an infinite normal form. But we have lost confluence of the finite $\lambda$-calculus. E.g.\ the finite term $(\lambda x.\Iterm (xx))(\lambda x.\Iterm (xx))$ has a finite reduction to $\Omegaterm$ and an infinite converging reduction to $\Iterm (\Iterm (\Iterm( \ldots)))$. Both reducts have the property that they can only reduce to themselves. Hence they cannot be joined by either finite or converging reductions. This example also shows that this extension of the finite lambda calculus is not normalising.
\label{infinitelambdacalculus}

Yet, it is possible to build  (in fact many different) infinitary extensions of $\lambda_\beta$ which are  confluent and normalising for finite and convergent reductions, and finite and infinite terms~\cite{KKSV97,KOV99,KV03,SV11}. We need to do three things. First,  we add a new symbol $\bot$ to the syntax of $\lambda$-terms and consider the set $\Lib$ of finite and infinite terms  over the extended coinductive syntax. Second, we choose a set $\setU$ of $\lambda$-terms in  $\Li$. Third,  we add
a new reduction $\bot_\setU$-rule on $\Lib$ that will
allow us to identify the terms of $\setU$  by the new symbol $\bot$:
\[\ProofRule{M [\bot:= \omegaterm]  \in \setU \ \ \ M \not = \bot}
{M \one \bot}{\bot_{\setU}}\]
For a given $\setU$ we denote this infinite extension by
$\libBU$.
In a series of papers~\cite{KKSV97,KOV99,KV03,SV11} we have determined a  collection of necessary and sufficient axioms that the set $\setU$ must satisfy in order for $\libBU$ to be a converging and normalising infinite $\lambda$-calculus. 

We call such sets {\em sets of meaningless terms}. The choice of a
set $\setU$ of meaningless terms is akin to the choice of a semantics for lambda
calculus: together the normal forms in $\libBU$ form a model of the
$\lambda$-calculus. The intuition is that the elements of a meaningless
set are undefined, that is, have no meaning or are insignificant. In order for
such a model to be consistent the set $\setU$ has to be a proper subset of
$\Li$.

\begin{defi}[\cite{SV11}]\label{meaninglesset}
 $\setU\subseteq \Li$ is called a {\em set of (finite or infinite)
 meaningless terms},
 if it satisfies the axioms of meaninglessness:
\begin{enumerate}
\item {\sf Axiom of  Root-activeness}: $\setroot \subseteq \setU$. ($\setroot$ defined below)
\item  
{\sf Axiom of Closure under $\beta$-reduction}: If $M \manybeta N$ then $N \in \setU$
for all $M \in \setU$.
\item  
 {\sf Axiom of  Closure under Substitution}:
 If   $M \in \setU$ then any substitution instance of $M$ is an element of $\setU$.

\item  
 {\sf Axiom of (Weak) Overlap}: Either for each  $\lambda x.P \in \setU$,
 there is some $W \in \setU$ such that
$P \manybeta W x\,$, or alternatively  $\,(\lambda x.P)Q \in \setU$, for any $Q \in \Lib$.

\item  
 {\sf Axiom of Indiscernibility}: Define  $M \indis N$ if $M$ can be transformed into $N$ by replacing  pairwise disjoint subterms of $M$ in $\setU$ by terms in $\setU$. If 
     $M \indis N$ then
         $M \in \setU \Leftrightarrow
 N \in \setU$.

 \item  {\sf Axiom of Consistency}: $\setU \neq \Lambda$.
\end{enumerate}
\end{defi}

This construction is inspired by the definition of B\"ohm
tree~\cite{Barendregt1984}. If one takes for $\setU$ the set $\setUnsolvables$
of unsolvables~\cite{bare:1975}, then the resulting infinite $\lambda$-calculus $\libBohm$
is confluent and normalising for $\beta{\bot_{\setUnsolvables}}$ reduction. The
B\"ohm tree  of a 
finite $\lambda$-term $M$ can equivalently be described as its
unique normal form in $\libBohm$~\cite{KKSV97}. Here a (possibly infinite) closed term $M$ in $\Li$ is called {\em solvable} if $M N_1\ldots N_k \finbeta
  \Iterm$ for some sequence $N_1,\ldots,N_k$ with $k\geq 0$. An open lambda
  term is called {\em solvable} if its closure is solvable. A $\lambda$-term is called {\em unsolvable} if it is not solvable.
  The set of unsolvables is the largest set for which this construction
  works. A $\lambda$-term is unsolvable if an only if it has no finite $\beta$-reduction to a
  head normal form~\cite{Barendregt1984}.

  The smallest set of meaningless terms~\cite{KKSV97,bera96} is the set $\RA$ of
terms that are root-active (or mute). A $\lambda$-term $M$ is
  root-active if any reduct of $M$ can further reduce to a redex. The
  classical root-active term is $\omegaterm$.  The unsolvable $\Omegaterm
  \Iterm$
  is not root-active. Note that the definition of a root-active
  term allows for free variables. The normal
forms in $\libbU{\RA}$ are exactly the Berarducci trees.

The L\'evy-Longo trees can be obtained if one performs this construction over
the set of terms without a weak head normal form. In general there are
uncountably many  sets of meaningless terms~\cite{SV11}. The collection of normal forms
of each such $\libBU$ is a model of the $\lambda$-calculus $\lb$. The axioms are
chosen such that different sets of meaningless terms give rise to different consistent models.

Church considered the terms without finite normal form as
insignificant~\cite{Church1941,Barendregt1984}.
We recognise that  the set of terms in $\Li$ without a
finite normal form is not a set of meaningless terms~\cite{KOV99,KV03}
in the sense of Definition~\ref{meaninglesset}, because it  is not
closed under reduction. The term $\Kterm \Iterm\Omegaterm$ has an infinite
reduction, because its subterm $\Omegaterm$ has. Yet $\Kterm\Iterm
\Omegaterm$ reduces to the finite normal form $\Iterm$. We will come back to this in Section~\ref{lambdaI}.


\section{Encoding many-valued logic in $\lambda$-calculus}
\label{sec3}
In this section we will extend the familiar Church encoding of Boolean logic to many-valued logic using ideas from B\"ohm trees and infinitary $\lambda$-calculus.
 We don't know  precise reference to the original Church encoding. As  Landin remarks in~\cite{Landin1964Mechanical}: \begin{quotation}In particular Church and Curry, and
McCarthy and the ALGOL 60 authors, are so large a
part of the history of their respective disciplines as to
make detailed attributions inevitably incomplete and
probably impertinent.\end{quotation} Berarducci and B\"ohm
have vastly generalised the Church encoding~\cite{BohmBerarducci1985}.

\subsection{Encoding Boolean logic in $\lambda$-calculus}\label{encodingBooleansd}
In ``the History of Lisp''~\cite{McCarthy1978} John McCarthy mentions his ``invention of the true
conditional expression $$\ITE{M}{N_1}{N_2}$$ which evaluates only one of $N_1$ and $N_2$
according to whether $M$ is true or false'' and also his ``desire for a
programming language that would allow its use'' in the period 1957-8. He also recalls  ``the conditional expression interpretation of
Boolean connectives'' as one of the characterising ideas of LISP. By this he
means concretely the if-then-else construct (when applied to Boolean
expressions only) which in combination with the truth values $\True$ and $\False$ can
be used as a basis for propositional logic~\cite{McCarthy1960} with the following natural definitions:
\begin{equation}\begin{array}{ccl}
  \neg   &\equiv & \lambda m .\ITE{m}{\False}{\True}\\
 \land  &\equiv & \lambda mn .\ITE{m}{n}{\False}\\
 \lor   &\equiv &  \lambda mn .\ITE{m}{\True}{n}\\
 \limp &\equiv &  \lambda mn .\ITE{m}{n}{\True}\\
  \end{array}\label{two}\end{equation}

Barendregt's book~\cite{Barendregt1984}  records 
two elegant encodings of the Booleans and the if-then-else construct. One encodes into the
classical 
$\lambda$-calculus and the other into the more restricted
$\lambda\Iterm$-calculus preferred by Church~\cite{Church1932,Church1941}. The latter we will discuss in Section~\ref{lambdaI}.
The former is the simplest: \label{lnot}
\[\begin{array}{lcl} \True&\equiv&\lambda x y . x\\
\False&\equiv&\lambda x y . y\\
 \ITE{B}{M}{N}&\equiv& B M N\\
\end{array}\]
It is easy to see that  if-then-else behaves as intended in this encoding. When $B$ reduces to $\True$ and $\False$,  we have respectively:
\begin{equation}\begin{array}{lcl}
\ITE{\True}{M}{N}&\finstep& M\\
 \ITE{\False}{M}{N}&\finstep& N
\end{array} \label{ITErules}\end{equation}
 With help of~(\ref{ITErules}) it is
straightforward to verify that the standard truth tables  of Figure~\ref{bool} for  Boolean
valued propositional logic hold in $\lambda$-calculus. 
\begin{figure}[htb]
\begin{tabular}{llll}
$\begin{array}{l|l}\lnot \\
\hline
\True &  \False\\
\False & \True\\
\end{array}$
 &
$\begin{array}{l|ll}\land&\True&\False\\
\hline
\True & \True & \False\\
\False & \False & \False\\
\end{array}$
&
$\begin{array}{l|ll}\lor&\True&\False\\
\hline
\True & \True & \True\\
\False & \True & \False\\
\end{array}$
&
$\begin{array}{l|ll}\rightarrow&\True&\False\\
\hline
\True & \True & \False\\
\False & \True & \True\\
\end{array}$\\
\end{tabular}
\caption{Boolean-valued  propositional logic\label{bool}}
\end{figure}
Boolean logic commonly deals with finite propositions. The set of  finite
propositions can be defined formally with an inductive syntax, where $p$ ranges over some possibly infinite set of propositional
variables: 
\begin{equation}\label{finProp}\phi ::= p 
\mid \True 
\mid \False
\mid (\phi \land \phi)
\mid (\phi \lor \phi)\mid (\phi \limp \phi)
\mid \neg \phi\end{equation}
It is not hard to prove by induction that all closed finite propositions have a
unique finite normal form:
\begin{lem} Let $\phi$ be a finite closed proposition. Then  $\phi$ has a unique finite normal form, which is either $\True$ or $\False$.
\end{lem}

\subsection{Encoding infinitary propositions in infinitary $\lambda$-Calculus}
Infinite propositions can be used to model certain while statements. For instance Bergstra and Ponse~\cite{BergstraPonse2011} model
 $\mathit{while\ } \neg a \mathit{\ test\ } b$ as the potentially infinite solution of the recursive equation\[W = \ITE{a}{\True}{ (\ITE{b}{W}{W})}\]
By reading the syntax definition~(\ref{finProp}) coinductively we obtain the
set of finite and infinite propositions.

In the introduction we showed how  Russell's paradox leads to  the infinite proposition $\neg (\neg (\neg (\ldots)))$. The encoding of
this infinite proposition in $\lambda$-calculus is the infinite term \[
  (((\ldots)\False\True)\False\True)\False\True\] which is an infinite normal
form with an
infinite left spine and no head normal form. Hence this encoding of the
Russel's paradox is unsolvable.

Not all infinite propositions reduce to infinite left spines: for instance, the infinite proposition
\[P_1 \equiv \True\land(\True \land \ldots)\equiv \True\land P_1 \equiv \True P_1 \True \equiv (\lambda x y . x) P_1 \True \to P_1\]
is root-active.
Also, some infinite propositions reduce just to  $\True$ or $\False$: for instance, the term
\[P_2 \equiv \True \lor (\True \lor (\True \lor (\ldots))) \equiv \True \lor P_2 \equiv  \True \True P_2   \equiv  (\lambda x y . x)\True  P_2   \to \True \]
These examples show that some infinite propositions reduce to a Boolean, but
not all. The latter have in common that their B\"ohm tree is $\bot$. 
\begin{thm}\label{Bohm} Let $\phi$ be a finite or infinite  closed proposition. Then  the B\"ohm tree
  of  
 $\phi$ is either $\True$, $\False$ or $\bot$.
\end{thm}
\begin{proof}By coinduction!\end{proof}
The missing detail in the above ``proof'' follows from the corollary of the next lemma:

\begin{lem}\label{presUnsolvables}
Let $U$ be an unsolvable $\lambda$-term in $\Lambda$. Then 
$\neg U$, $ U \land N$, $U \lor N$ and $ U \limp N$ are all unsolvable terms.
\end{lem}\begin{proof}Immediate from the definitions. For instance, suppose
  $\neg U$ is solvable; then \[(\neg U) N_1 \ldots N_n \fin \Iterm\] for some $ N_1,
  \ldots, N_n$. But $\neg U \equiv U \False \True$. Hence $U$ is
  solvable.  Therefore,  unsolvability of $U$ implies the  unsolvability of $\neg U$.
\end{proof}

\begin{cor}\label{presUnsolvables2}
 The B\"ohm trees of\/
$\neg \bot$, $ \bot \land N$, $\bot \lor N$ and $ \bot \limp N$ are all equal to $\bot$.
\end{cor}
\begin{proof} Eg., $B(\neg \bot) = B(\neg U)= B(U \False \True) = \bot$
\end{proof}

\subsection{Encoding three-valued McCarthy logic with help of  B\"ohm trees}
 
Theorem~\ref{Bohm} suggests an experiment: what logic do we obtain if  we
repeat the encoding of Section~\ref{encodingBooleansd} with three truth values
$\{\True,\False,\bot\}$ instead of two?
Using Lemma~\ref{presUnsolvables2}  we can extend the truth tables of Boolean-valued logic to  the truth tables of
Figure~\ref{fivefive}. These are exactly the  truth tables of McCarthy's left-sequential three-valued
 propositional logic. We will use the notation of~\cite{BergstraBethkeRodenburg1995} and write $\lsand$ and $\lsor$ for the conjunction and disjunction in left-sequential logic.
\begin{figure}[t]
\begin{tabular}{llll}
$\begin{array}{l|l}\lnot\\
\hline
\True &  \False\\
\False & \True\\
\bot & \bot
\end{array}$
 &
$\begin{array}{l|lll}\lsand&\True&\False&\bot\\
\hline
\True  & \True  & \False & \bot \\
\False & \False & \False & \False \\
\bot & \bot & \bot & \bot \\
\end{array}$
&
$\begin{array}{l|lll}\lsor&\True&\False&\bot\\
\hline
\True  & \True  & \True & \True\\
\False & \True  & \False & \bot\\
\bot & \bot & \bot & \bot \\
\end{array}$
&
$\begin{array}{l|lll}\rightarrow&\True&\False&\bot\\
\hline
\True  & \True  & \False & \bot \\
\False & \True & \True & \True \\
\bot & \bot & \bot & \bot \\
\end{array}$
\end{tabular}
\caption{McCarthy's   left-sequential three-valued propositional logic\label{fivefive}}
\end{figure}

McCarthy discovered  left-sequential three-valued propositional logic in his search for a suitable formalism for a
mathematical theory of computation~\cite{McCarthy1963}. In the context of a language for
computational (partial) functions  he introduced conditional expressions of the form\[(p_1 \to e_1,\ldots,p_n \to e_n)\]
where the $p_i$ are propositional expressions that evaluate to true or false. The idea is that
 the value of the whole conditional expression is the value of the
expression $e_i$  for the first $p_i$ with value true. If all $p_i$
have value false then the conditional expression is undefined. To allow that
 the evaluation of an expression can be inconclusive, McCarthy stated the rule to evaluate  conditional expressions more precisely:
\begin{quote} If an undefined $p$ occurs before a true $p$ or if all $p$'s are false or if the $e$ corresponding to the first true $p$ is undefined, then the form is undefined. Otherwise, the value of the form is the value of the $e$ corresponding to the first true $p$. \end{quote}
Now the propositional connectives can be defined with help of conditional expressions.
\begin{equation}\begin{array}{ccl}
  \neg  p &\equiv & (p\to \False,\True \to \True)\\
p \lsand q &\equiv & (p\to q,\True \to \False)\\
p \lsor  q &\equiv & (p\to \True,\True \to q)\\
p \limp q &\equiv & (p\to q,\True \to \True)\\
\end{array}\label{three}\end{equation}
for which McCarthy then derives the very same truth tables of Figure~\ref{fivefive}. In the presence of the third truth value, undefined, the left sequential conjunction and disjunction are no longer commutative. 

Guzman and Squier~\cite{GuzmanConditionalLogic} gave   a complete axiomatisation of McCarthy's logic,
cf.\ Figure~\ref{completeAx}. They also gave the following definition of conditional using the left-sequential connectives.
\begin{figure}[ht]$\begin{array}{|lrcl|}\hline&&&\\
(1) &\lnot \True &=&  \False  \\
(2) &\lnot \bot &=& \bot \\
(3) &\lnot\lnot x &=& x  \\
(4) &\lnot(x \lsand y) &=& \lnot x \lsor \lnot y\\
(5) &x \to y &=& \lnot x \lsor y   \\
(6) &x \lsand (y  \lsand z) &=& (x \lsand y)  \lsand z  \\
(7) &\True \lsand x &=&  x \\
(8) &x \lsor (x \lsand y) &=& x  \\
(9) &x \lsand (y \lsor z) &=& (x \lsand y)\lsor( x \lsand z)  \\
(10) &(x \lsor y) \lsand z &=& (x \lsand z)  \lsor (\lnot x \lsand y\lsand z) \\
(11) & (x \lsand y)\lsor( y \lsand x)  &=&  (y \lsand x)\lsor( x \lsand y) \\ &&&\\\hline 
  \end{array}
 $
  \caption{\label{sixsix}Complete axiomatisation of McCarthy's
  left-sequential three-valued
  propositional logic by Guzman and Squire\label{completeAx}}
\end{figure}
\begin{lem} $\ITE{B_0}{B_1}{B_2}=(B_0\lsand B_1)\lsor(\neg  B_0 \lsand B_2)$ for all $B_0,B_1,B_2\in\{\True, \bot,\False\}$. \end{lem}
\begin{proof} After applying the definitions of the logical operators it remains to show that
  \begin{equation}B_0 B_1 B_2= (B_0 B_1 \False) \True ((B_0\False\True) B_2\False)\label{seven}\end{equation}
The argument now is by inspection.
  \begin{itemize}
  \item $B_0=\True$. Since $\True x y = x$,  it is sufficient to show that
     $ B_1 = B_1 \True \False$. This follows by inspection of the three options for $B_1\in \{\True, \bot,\False\}$. 
  \item $B_0=\bot$. Then $\bot B_1 \bot= \bot =  (\bot B_1 \False) \True ((\bot\False\True) B_2\False)$.
  \item $B_0=\False$.   Since $\False x y = y$, it is enough to show that $B_2 = \True  B_2 \False$, which follows by   $\True x y = x$.
  \end{itemize}
  \end{proof}
In the remainder  of the paper we will ignore  implication as it can be defined from $\lnot$ and $\lsor$.
\subsection{Refining the encoding from three-valued to four- and five-valued logic}\label{refineToFourAndFive}

In the previous section we identified unsolvable $\lambda$-terms with 
 $\bot$, their
 (possibly infinite)  normal form  in the infinitary $\lambda$-calculus
$\libBohm$. We used $\bot$ as third truth value besides $\True$ and $\False$.
We can refine this idea using the observation of~\cite{SVwollic11} that the set of
unsolvables is the union of three pairwise disjoint sets, each closed under substitution and 
infinite reduction.

At the basis of this observation lies the simple and well known fact that any {\em finite} $\lambda$-term has one of two forms, where $m,n$ range over natural numbers:
\[\begin{array}{l}\lambda x_1\ldots x_n . x M_m \ldots M_1\\ 
\lambda x_1\ldots x_n . (\lambda x P) Q M_m \ldots M_1\end{array}\]
The former expression is called a {\em head normal form} and the redex $(\lambda x P) Q$ in the latter is called the head redex. Wadsworth has shown that repeated head reduction of a term $M$ terminates in a head normal form if and only if  $M$ has one, and also that having a head normal form is equivalent to being solvable. A reduction in which each step reduces a head redex is called a head reduction~\cite[see Section 8.3]{Barendregt1984}

Hence any unsolvable term $M$  has an infinite head reduction. One of the following scenarios must hold for $M$.
\begin{itemize} \item  $M$ has an infinite head reduction to a term of the form $\lambda x \lambda x \lambda x \ldots$ (modulo renaming). Example $\Yterm \Kterm \manystep\lambda x \lambda x \lambda x \ldots$.
\item $M$ has an infinite head reduction to a term of the form $ \lambda x_1 \ldots\lambda x_n.((((\ldots) M_3)M_2)M_1)$. An example $\Yterm \lambda x.xy \manystep (((\ldots y)y)y)$.
  \item $M$ has a finite head reduction to a term of the form $\lambda x_1\ldots x_n . (\lambda x P) Q M_m \ldots M_1$ in which the head redex $(\lambda x.P)Q$ is root active. The term $\Omegaterm$ is an example.
\end{itemize}

These three mutually exclusive fates of unsolvable terms lead to the following definition.
\begin{defi}[\cite{SVwollic11}]
\begin{enumerate}
\item   $\sethr = \{ M \in \Linob \mid M \finbeta N 
                      \mbox{ and $N$ is a head active form} \}$
where 
$N$ is a {\em head  active form}  
if  $M = \lambda x_1 \ldots x_n. R P_1 \ldots P_k$ and $R$ is root-active.
\item  
  $\setil = 
  \{ M \in \Linob \mid M \manybeta N \mbox{ and $N$ is an infinite left
  spine form} \}$
where
$N$ is an  {\em infinite left spine form} 
if $N = \lambda x_1 \ldots x_n. (( \ldots P_2)P_1$.
\item
 $\setogre = \{ M \in \Linob \mid M \manybeta \Oterm \}$.
\end{enumerate}\end{defi}

The three sets can be characterised alternatively using the notion of Berarducci
tree which can reveal more detail of a term than B\"ohm trees do.
\begin{lem}
\begin{enumerate}
\item   $ M\in \sethr$ if and only if the Berarducci tree of $M$ is of the form
\[\lambda x_1\ldots  x_n.\bot N_m\ldots N_1\] for some natural numbers $n,m$.

\item  $ M\in \setil$ if and only if the Berarducci tree of $M$ is of the form
\[\lambda x_1\ldots x_n.((((\ldots) N_3)N_2) N_1)\]for some natural numbers $n$.
\item
 $M \in \setogre$ if and only if  the Berarducci tree of $M$ is $\lambda x_1 x_2
 x_3\ldots$, ie.\ $\Oterm$.

\end{enumerate}\label{refine}\end{lem}

The union of the $\sethr$,  $\setil$ and $\setogre$ is the set of unsolvables.
%
With help of these three sets we can
refine the notion of B\"ohm reduction. We will represent each set by its own truth value.
Instead of replacing unsolvable all $\lambda$-terms by $\bot$ we will now replace the
elements in $\setHA$, $\setIL$ and $\setO$ by, respectively, the constants $\botHA$, $\botIL$ and 
$\botO$, so that instead of one $\bot$-reduction $\to_\bot$ we have now three
reduction rules, that we denote by $\to_{\botHA}$, $\to_{\botIL}$ and 
$\to_{\botO}$. We will use $\botHA$, $\botIL$ and 
$\botO$ as truth values next to $\True$ and $\False$ to interpret five-valued  propositional logic.

In the same fashion, if we split the unsolvables in only two sets $\setHA$ and $\setIL\cup\setO$ and introduce besides $\botHA$ a single constant $\botILO$ to replace the
elements in $\setIL\cup\setO$,  we  have the ingredients to interpret four-valued  propositional logic.

These constructions work because of the following theorem.
\begin{thm}
  \begin{enumerate} \item Let $\Lambda^\infty_{\botHA\botIL\botO}$ be the set of
  finite and infinite $\lambda$-terms constructed with the symbols $\botHA$, $\botIL$ and 
  $\botO$. Then the infinitary $\lambda$-calculus
  $\libbUUU{\setHA}{\setIL}{\setO}$ is
  confluent and normalising for (strongly) convergent reduction.

\item Let $\Lambda^\infty_{\botHA\botILO}$ be the set of
  finite and infinite $\lambda$-terms constructed with the symbols $\botHA$ and $\botILO$. Then the infinitary $\lambda$-calculus $\libbUU{\setHA}{\setIL\cup\setO}$ is
  confluent and normalising for (strongly) convergent reduction.
  \end{enumerate}
\end{thm}
\begin{proof}
Both follow from Lemma~\ref{refine} and two facts from~\cite{KKSV97}, namely that $\libbU{\setR}$ is confluent and normalising,
and that $\bot_\setR$-reduction can be postponed over $\beta$-reduction.
\end{proof}

We will now encode five-valued logic in $\lambda$-calculus using the same logical operators as
before together with the five truth values from $\{\True,\False,
\botHA,\botIL,\botO\}$. Similarly using the four truth values from $\{\True,\False,
\botHA,\botILO\}$ we will encode four-valued logic.

We need an analogue of Corollary~\ref{presUnsolvables2}.

\begin{lem}\label{threeten}
Let $U$ be a  $\lambda$-term in $\setHA$ ($\setIL$, $\setO$ and $\setIL\cup\setO$ ). Then 
$\neg U$, $ U \lsand N$, $U \lsor N$ and $ U \limp N$ are all terms in  $\setHA$ ($\setIL$, $\setO$ and $\setIL\cup\setO$).
\end{lem}\begin{proof}Immediate from the definitions. For instance, suppose $U \in \setHA$, that is suppose the
  Berarducci tree of $U$ is of the form $\lambda x_1\ldots  x_n.\bot N_m\ldots
  N_1$. Then the Berarducci tree of $\neg U$ is the  Berarducci tree of $(\lambda x_1\ldots  x_n.\bot N_m\ldots
  N_1)\False\True$. One easily sees that $\neg U$ is an element of $\setHA$.
\end{proof}

\begin{cor}\label{HAIlO}
  \begin{enumerate}\item
 The normal forms  in  $\libbUUU{\setHA}{\setIL}{\setO}$ of
$\neg \bot_\mathcal{X}$, $ \bot_\mathcal{X} \lsand N$, $\bot_\mathcal{X} \lsor N$ and $
 \bot_\mathcal{X} \limp N$ are all equal to $\bot_\mathcal{X}$ for $\mathcal{X} \in \{\setHA,\setIL,\setO\}$.
\item
 The normal forms  in  $\libbUU{\setHA}{\setIL\cup\setO}$ of
$\neg \bot_\mathcal{X}$, $ \bot_\mathcal{X} \lsand N$, $\bot_\mathcal{X} \lsor N$ and $
 \bot_\mathcal{X} \limp N$ are all equal to $\bot_\mathcal{X}$ for $\mathcal{X} \in \{\setHA,\setIL\cup\setO\}$.
  \end{enumerate}
\end{cor}

\begin{thm}
  \begin{enumerate}
\item  Let $\phi$ be a finite or infinite  closed proposition with truth values from $\{\True,\False, \botHA, \botIL, \botO\}$. Then  the  normal form  of  $\phi$ in  $\libbUU{\setHA}{\setIL}{\setO}$  is either $\True$, $\False$,
  $\botHA$,  $\bot_{\setIL}$ or $\bot_\setO$.
\item  Let $\phi$ be a finite or infinite  closed proposition with truth values from  $\{\True,\False, \botHA, \botILO\}$. Then  the  normal form   of  $\phi$ in  $\libbUU{\setHA}{\setIL\cup\setO}$  is either $\True$, $\False$,
  $\botHA$ or  $\botILO$.
   \end{enumerate}
\end{thm}
\begin{proof}By coinduction!\end{proof}

Using Corollary~\ref{HAIlO}, it is straightforward to calculate  the truth tables for a four-valued logic encoded in $\lambda$-calculus:

\begin{figure}[h]\begin{tabular}{c}
\begin{tabular}{cc}
$\begin{array}{l|l}\lnot\\
\hline
\True &  \False\\
\False & \True\\
\botHA & \botHA\\
\botILO & \botILO\\
\end{array}$
 &\quad
$\begin{array}{l|lllll}\lsand&\True&\False& \botHA & \botILO\\
\hline
\True  & \True  & \False & \botHA & \botIL \\
\False & \False & \False & \False     & \False    \\
\botHA & \botHA & \botHA & \botHA & \botHA\\
\botILO & \botILO & \botILO & \botILO & \botILO \\

\end{array}$
\end{tabular}
\\\\
$\begin{array}{l|lllll}\lsor&\True&\False&\botHA & \botILO \\
\hline
\True  & \True  & \True & \True & \True\\
\False & \True  & \False & \botHA & \botILO \\
\botHA & \botHA & \botHA & \botHA & \botHA  \\
\botILO & \botILO & \botILO & \botILO & \botILO \\
\end{array}$
\end{tabular}
\caption{Left-sequential four-valued propositional logic\label{fourValuedLogic}}
\end{figure}

As it happens, this four-valued propositional logic has been
studied by Bergstra and Van de
Pol~\cite{BergstraPol1996,BergstraPol2011}.  In the context of process algebra enriched with conditional statements the
need for many-valued logic arises in case a condition evaluates to a
truth value (\eg, error/exceptions and divergences) different from true
or false. This led Bergstra and his colleagues to a study
of a great many of versions of three-, four- and even five-valued
logic~\cite{BergstraBethkeRodenburg1995,BergstraPonse1998BochvarMcCarthy,BergstraPonse2000,BergstraPonse1999}.

For the four-valued logic of Figure~\ref{fourValuedLogic} Bergstra and Van de Pol gave a complete axiomatisation  in~\cite{BergstraPol1996,BergstraPol2011}. See Figure~\ref{AxiomsfourValuedLogic}. They use $\mathbf{m}$ (meaningless) for $\botHA$ and $\mathbf{d}$ (divergence) for $\botILO$. These names make some sense here as well. The terms in $\setIL\cup\setO$ can be called diverging as they have limits with infinite left spines.  On the other hand, terms in $\setHA$ reduce by definition to terms of the form  $M = \lambda x_1 \ldots x_n. R P_1 \ldots P_k$ with $R$ is root-active. This term $R$ is meaningless, in the sense that it will not reveal any further information how long one may reduce it.

\begin{figure}[h]\begin{tabular}{c}
\begin{tabular}{cc}
$\begin{array}{|lrcl|}\hline &&&\\
(1)&\lnot \mathbf{d} &=&  \mathbf{d}  \\
(2)&\lnot \mathbf{m} &=&  \mathbf{m}  \\
(3)&\lnot \True &=&  \False  \\
(4)&\lnot\lnot x &=& x  \\
(5)&\True \lsand x &=&  x \\
(6)&\False \lsand x &=&  \False \\
(7)&x \lsor y &=& \lnot(\lnot x \lsand \lnot y)\\
(8)&x \lsand (y  \lsand z) &=& (x \lsand y)  \lsand z  \\
(9)&(x \lsor y) \lsand z &=& (\lnot x \lsand y\lsand z)\lor (x \lsand z)\lsor (x \lsand z) \\ && & \\
\hline 
\end{array}$
\end{tabular}\end{tabular}
\caption{Bergstra and Van de Pol's axiomatisation of left-sequential four-valued propositional logic\label{AxiomsfourValuedLogic}}
\end{figure}

The axioms in Figure~\ref{AxiomsfourValuedLogic} have been selected carefully: each is independent of the others. They also note that Axiom (11) of Figure~\ref{sixsix},  \[(x \lsand y)\lsor( y \lsand x)  =  (y \lsand x)\lsor( x \lsand y)\]
does not hold in four-valued logic. We can recognise that in our context if we substitute $\botHA$ for $x$ and $\bot_{\setIL\cup\setO}$ for $y$. Then by Lemma~\ref{threeten} we see immediately that the left-hand side of the axiom reduces to  $\botHA$, while the right-hand side reduces to $\bot_{\setIL\cup\setO}$.

Similarly, the truth tables for the five-
valued logic encoded in $\lambda$-calculus are as follows:

\begin{figure}[th]\begin{tabular}{c}
\begin{tabular}{cc}
$\begin{array}{l|l}\lnot\\
\hline
\True &  \False\\
\False & \True\\
\botHA & \botHA\\
\botIL & \botIL\\
\botO & \botO\quad\\
\end{array}$
 &\quad
$\begin{array}{l|lllll}\lsand&\True&\False& \botHA & \botIL &\botO\\
\hline
\True  & \True  & \False & \botHA & \botIL & \botO\\
\False & \False & \False & \False     & \False    & \False\\
\botHA & \botHA & \botHA & \botHA & \botHA & \botHA \\
\botIL & \botIL & \botIL & \botIL & \botIL & \botIL \\
\botO & \botO & \botO & \botO & \botO & \botO\\
\end{array}$
\end{tabular}
\\\\
$\begin{array}{l|lllll}\lsor&\True&\False&\botHA & \botIL &\botO\\
\hline
\True  & \True  & \True & \True & \True& \True\\
\False & \True  & \False & \botHA & \botIL & \botO\\
\botHA & \botHA & \botHA & \botHA & \botHA & \botHA \\
\botIL & \botIL & \botIL & \botIL & \botIL & \botIL \\
\botO & \botO & \botO & \botO & \botO & \botO\\
\end{array}$
\end{tabular}
\caption{Left-sequential five-valued propositional logic\label{fiveValuedLogic}}
\end{figure}

Finally using Corollary~\ref{HAIlO}, it is also straightforward to calculate  the truth tables of a five-valued logic encoded in $\lambda$-calculus. See Figure~\ref{fiveValuedLogic}.  This the five-valued logic that Bergstra and Van de Pol left implicit in their final remark in ~\cite{BergstraPol2011} that their complete axiomatisation generalises to five- and higher valued logics, as  long as one adds axioms of the form $\lnot p = p$ for each new truth value $p$.

\section{Encoding three-valued logic in  the finite $\lambda\Iterm$-calculus}
\label{sec4}
The $\lambda$-calculus that Church used in his unfortunate attempt towards  a foundation of
mathematics was the $\lambda\Iterm$-calculus. This calculus differs from the common $\lambda$-calculus $\lb$ by a restriction on the set of
$\lambda$-terms.  Terms in the $\lambda\Iterm$-calculus only contain
abstractions of the form $\lambda x.M$ if $x$ occurs free in $M$.  
For example the terms $\lambda xy.x$ and $\lambda xy.y$ that we used for the
Booleans are now forbidden. So we cannot use the Church encoding of Boolean
logic as before. 

The consequence of this restriction is that terms in the
$\lambda\Iterm$-calculus have two properties Church deemed important: (i) if a term has a
finite normal form, it cannot have an infinite reduction, and (ii)
 if a term has a finite normal form then all its subterms  must also have a normal
form~\cite{Church1941}. These properties don't hold in the classical $\lambda$-calculus.

\subsection{Another encoding of the Booleans}
Barendregt gave in fact two  encodings for the Booleans in his
book~\cite{Barendregt1984}. Besides the previous well-known encoding of the Booleans he
also defined an encoding of the Booleans in the spirit of Church, because the new encodings of the Booleans are
terms in the $\lambda\Iterm$-calculus.
\[\begin{array}{lll}
\True_\Iterm & =& \lambda xy. y\Iterm\Iterm x\\
\False_\Iterm &=& \lambda x. x\Iterm\Iterm\Iterm 
\end{array}\]
In this case we cannot derive~(\ref{ITErules}). Instead we get
 \begin{equation}\begin{array}{lcl}
  \ITE{\True_\Iterm}{M}{N}&\finstep& N \Iterm\Iterm M \\
  \ITE{\False_\Iterm}{M}{N}&\finstep& M  \Iterm\Iterm\Iterm N
 \end{array}\label{threethree}\end{equation}

 Yet by inspection of each of the four concrete options for $M,N \in \{\True_\Iterm,\False_\Iterm\}$  we find that  \begin{equation}\begin{array}{lcl}
  N \Iterm\Iterm M  &\finstep& M\\
 M  \Iterm\Iterm\Iterm N  &\finstep& N
 \end{array}\label{fourfour}\end{equation}
Combining~(\ref{threethree}) with~(\ref{fourfour}) gives us~(\ref{ITErules})
for all Booleans  $M,N \in \{\True_\Iterm,\False_\Iterm\}$.  Hence also this less well-known encoding validates the truth tables of
Boolean propositional logic.

\subsection{B\"ohm trees in the finite $\lambda\Iterm$-calculus}\label{lambdaI}
Church strongly preferred the $\lambda\Iterm$-calculus over the unrestricted
$\lambda$-calculus.
For him the natural notion of the meaning of a $\lambda$-term is its finite
normal form, provided it exists. Terms without  finite normal form he considered
to be meaningless or, in his own alternative wording,  insignificant~\cite{Church1941}.
In the $\lambda\Iterm$-calculus terms without finite normal form can
safely be identified. In the unrestricted calculus this leads to
inconsistency~\cite[Proposition 2.2.4]{Barendregt1984}.

In fact, Barendregt~\cite{Barendregt1973lambdaI} has
shown that the unsolvable terms in the
$\lambda\Iterm$-calculus  are precisely the terms without finite normal
form. Klop~\cite{Klop1975} gave a simpler proof. They did not consider the B\"ohm tree construction. But in setting of the
$\lambda\Iterm$-calculus the B\"ohm tree construction  simplifies
enormously. There is no need to
consider infinite terms and infinite reductions. We just add the fresh symbol $\bot$ to the syntax of the $\lambda\Iterm$-calculus plus the rule\[ M
\to_\bot \bot\mbox{, whenever\ } M[\bot:=\Omegaterm] \mbox{\ has no finite normal form.}\] 
Let us denote this extension of the 
$\lambda\Iterm$-calculus by
$\lambda\Iterm_{\beta\bot}$. The extension $\lambda\Iterm_{\beta\bot}$ is confluent and
normalising in the finitary sense, and the B\"ohm tree of any $\lambda\Iterm$-term equals either $\bot$ or is a
finite $\bot$-free normal form. In the past we have overlooked this
construction, after observing that the set of $\lambda\Iterm$-terms is not closed under infinite
$\beta$-reduction. In the limit a bound variable may ``drop off''.
For instance, consider $\lambda v . \Yterm
(\lambda xyz.xy) v$. Since
\[\lambda v . \Yterm (\lambda xyz.xy) v \to
  \lambda v . (\lambda xyz.xy) (\Yterm (\lambda xyz.xy)) v \fin
  \lambda v \lambda z. \Yterm (\lambda xyz.xy) v
\]
we find that  the finite $\lambda\Iterm$-term $\lambda v . \Yterm
(\lambda xyz.xy) v$ has an infinite reduction to the infinite term $\lambda v \lambda
  z\lambda z\lambda z \ldots$ which is no longer a  $\lambda\Iterm$-term.

The above $\bot$-rule resolves this problem, because now the term $\lambda v . \Yterm
(\lambda xyz.xy) v$ reduces in one step to $\bot$. In $\lambda\Iterm_{\beta\bot}$ there is no need to consider infinite reduction as any finite  $\lambda\Iterm$-term has a finite reduction to finite B\"ohm tree in  $\lambda\Iterm_{\beta\bot}$.
Hence we can encode three valued-logic in  $\lambda\Iterm_{\beta\bot}$  if we
take as truth values  $\True_\Iterm,\False_\Iterm$ and $\bot$.

\begin{lem}\label{presUnsolvablesI}
Let $U$ be an unsolvable $\lambda\Iterm$-term. Then 
$\neg U$, $ U \lsand N$, $U \lsor N$ and $ U \limp N$ are all unsolvable
$\lambda\Iterm$-terms, when $N$ is a $\lambda\Iterm$-term.
\end{lem}\begin{proof}By Lemma~\ref{presUnsolvables} it  remains to show
    that $\neg U$, $ U \lsand N$, $U \lsor N$ and $ U \limp N$ are
    $\lambda\Iterm$-terms. But this follows from the fact that $\True_\Iterm$
    and $\False_\Iterm$ are $\lambda\Iterm$-terms.
\end{proof}
Thus, despite a different encoding of the Booleans, we
find  the same truth tables of  McCarthy's left-sequential  three-valued
propositional logic of Figure~\ref{fivefive}.
Note that the earlier partition in Section~\ref{refineToFourAndFive} of the unsolvables based on the form of the left spine
of their Berarducci tree applies verbatim to $\lambda\Iterm$-terms. Hence, also this second encoding of the
Booleans refines to an encoding of the same earlier four- and five-valued logics in the  $\lambda\Iterm$-calculus.

\subsection{Why Curry's Paradox does not  apply}
We end with noting that Curry's Paradox does not apply to the finite $\lambda\Iterm$-calculus because the above infinitary extension is consistent.

Contemporaneously with Church, Curry had been searching   for a symbolic logic that could serve as foundation of
mathematics. The technique by which Kleene and Rosser~\cite{KleeneRosser1935} found the inconsistency in the symbolic logic of Church also applied to some of the systems of illative combinatoric
logic that Curry was exploring. In contrast to Church, Curry had not committed
himself to an underlying philosophy. He considered the  Kleene-Rosser paradox an helpful instrument in the search for ``stronger and stronger systems which are consistent'' as well as ``weaker and weaker systems which are inconsistent''~\cite{CurryFeys1958CombinatoryLogic}.

In 1942 Curry
published a short and self-contained argument to show the inconsistency for the
type of symbolic logics that he and Church were working on. Curry showed that that any combinatory complete system, like e.g.\ $\lambda$-calculus, with
an implication operator satisfying:
\[X \rightarrow X\]
\[(X \to (X\to Y)) \to(X\to  Y)\]
is inconsistent. The elegant short proof of the Curry's Paradox can be found in~\cite{CurryFeys1958CombinatoryLogic,Barendregt1984}.

As the  infinitary extensions $\lb$ and the $\lambda\Iterm$-calculus$\lambda$-calculus are consistent (the normal forms of $\True$ and $\False$ are not equal in them) the Curry's Paradox does not apply to them. More direct: the implication $X \to Y$ does not satisfy the above two conditions for implication: if $X$ is $\bot$ then both expressions reduce to $\bot$ for any value of $Y$.

\section{Conclusion}
\label{sec5}
The idea to solve Russell's paradox with three-valued logic is not at
all new. Feferman gave various pointers
in~\cite{Feferman1984}. But the  conjunctions and
disjunctions  of the three-valued logics that are
considered for that purpose  all seem to be commutative in contrast to those in the left-sequential McCarthy
logic that we use here. 

It is possible to further refine the encoding  to an encoding of
$\infty$-valued logic  in $\lambda$-calculus. The new
truth values then correspond to the different shapes of left spine
that unsolvables can have. We see no further use for that.
\section*{Acknowledgements}
We would like to thank the editor for her encouragement and infinite patience, and the referees for careful reading of our paper and their constructive comments that helped to improve this paper.
\bibliography{newreferences}

\newcommand{\botomega}[2]{\bot^{#1}_{#2}}
\end{document}